\pdfoutput=1

\documentclass[11pt,twoside,a4paper,cmspaper,final,collab]{cms-tdr}
\def\svnVersion{427239P}\def\cmsCernNoTag{CERN-EP-2016-042}\def\cmsCernDate{\today}\def\cmsMessage{Published in Physical Review C as \href{http://dx.doi.org/10.1103/PhysRevC.96.014915}{\doi{10.1103/PhysRevC.96.014915.}}}

\begin{document}\cmsNoteHeader{HIN-14-008}

\hyphenation{had-ron-i-za-tion}
\hyphenation{cal-or-i-me-ter}
\hyphenation{de-vices}
\RCS$Revision: 427126 $
\RCS$HeadURL: svn+ssh://svn.cern.ch/reps/tdr2/papers/HIN-14-008/trunk/HIN-14-008.tex $
\RCS$Id: HIN-14-008.tex 427126 2017-09-28 06:41:44Z velkovsk $
\newlength\cmsFigWidth
\ifthenelse{\boolean{cms@external}}{\setlength\cmsFigWidth{0.85\columnwidth}}{\setlength\cmsFigWidth{0.4\textwidth}}
\newlength{\ruleht}
\setlength{\ruleht}{\baselineskip}
\ifthenelse{\boolean{cms@external}}{\providecommand{\breakhere}{\relax}}{\providecommand{\breakhere}{\linebreak[4]}}
\ifthenelse{\boolean{cms@external}}{\providecommand{\herebreak}{\linebreak[4]}}{\providecommand{\herebreak}{\relax}}
\ifthenelse{\boolean{cms@external}}{\providecommand{\cmsLeft}{top\xspace}}{\providecommand{\cmsLeft}{left\xspace}}
\ifthenelse{\boolean{cms@external}}{\providecommand{\cmsRight}{bottom\xspace}}{\providecommand{\cmsRight}{right\xspace}}
\newcommand {\rootsNN}  {\ensuremath{\sqrt{\smash[b]{s_{_{NN}}}}}\xspace}
\newcommand {\deta}     {\ensuremath{\Delta\eta}\xspace}
\newcommand {\dphi}     {\ensuremath{\Delta\phi}\xspace}
\newcommand {\pttrg}       {\ensuremath{p_\mathrm{T}^{\text{trig}}}\xspace}
\newcommand {\ptass}       {\ensuremath{p_\mathrm{T}^{\text{assoc}}}\xspace}
\newcommand {\etatrg}       {\ensuremath{\eta_\text{lab}^{\text{trig}}}\xspace}
\newcommand {\etaassoc}       {\ensuremath{\eta_\text{lab}^{\text{assoc}}}\xspace}
\newcommand {\etacm}       {\ensuremath{\eta_{\text{c.m.}}}\xspace}
\newcommand {\etalab}       {\ensuremath{\eta_{\text{lab}}}\xspace}
\newcommand {\pPb}  {\ensuremath{\mathrm{pPb}}\xspace}
\newcommand {\pp}  {\ensuremath{\mathrm{pp}}\xspace}
\newcommand {\pA}    {\ensuremath{\mathrm{p}\mathrm{A}}\xspace}
\newcommand{\noff}    {\ensuremath{N_\text{trk}^\text{offline}}\xspace}
\ifthenelse{\boolean{cms@external}}{\providecommand{\NA}{\ensuremath{\cdots}}\xspace}{\providecommand{\NA}{---\xspace}}
\newcolumntype{x}{D{,}{\,\pm\,}{4.8}}
\cmsNoteHeader{HIN-14-008}
\title{Pseudorapidity dependence of long-range two-particle correlations in pPb collisions at \texorpdfstring{$\sqrt{s_{_{NN}}}=5.02\TeV$}{sqrt(s[NN])=5.02 TeV}}

\date{\today}

\abstract{
Two-particle correlations in \pPb collisions at a nucleon-nucleon center-of-mass energy
of 5.02\TeV are studied as a function of the pseudorapidity separation ($\Delta \eta$) of the particle pair at small relative azimuthal angle ($\abs{\Delta \phi}< \pi/3$). The correlations
are decomposed into a jet component that dominates the short-range correlations ($\abs{\Delta \eta} < 1$), and a component that
persists at large $\Delta \eta$ and may originate from collective behavior of the produced system. The events are classified in terms of the multiplicity of the produced particles. Finite azimuthal anisotropies are observed in high-multiplicity events. The second and third Fourier components of the particle-pair azimuthal correlations, $V_2$ and $V_3$, are extracted after subtraction of the jet component.
The single-particle anisotropy parameters $v_2$ and $v_3$ are normalized by their lab frame mid-rapidity value and are studied as a function of $\etacm$. The normalized $v_2$ distribution is found to be asymmetric about $\etacm = 0$, with smaller values observed at forward pseudorapidity, corresponding to the direction of the proton beam, while
no significant pseudorapidity dependence is observed for the normalized $v_3$ distribution within
the statistical uncertainties.}

\hypersetup{%
pdfauthor={CMS Collaboration},%
pdftitle={Pseudorapidity dependence of long-range two-particle correlations in pPb collisions at sqrt(s[NN])=5.02 TeV},%
pdfsubject={CMS},%
pdfkeywords={CMS, physics, heavy ion,two-particle long-range correlations}}

\maketitle

\section{Introduction}
\label{sec:intro}
Studies of two-particle correlations play an important role in understanding the underlying mechanism of particle production in high-energy nuclear collisions ~\cite{Adams:2005dq,Adcox:2004mh,Back:2004je}.  Typically, these correlations are studied in a two-dimensional $\dphi$-$\deta$ space, where $\dphi$ and $\deta$ are the differences in the azimuthal angle $\phi$ and the pseudorapidity $\eta$ of the two particles.

A notable feature in the two-particle correlations is the so-called ``ridge,'' which is an extended correlation structure in
relative pseudorapidity $\deta$ concentrated at small relative azimuthal angle $|\dphi| \approx 0$. The ridge, first observed
in nucleus-nucleus (AA) collisions~\cite{Adams:2005ph,Alver:2009id,Abelev:2009af}, has been studied both at RHIC and LHC over a
wide range of collision energies and system sizes~\cite{Adams:2005ph,Alver:2009id,Abelev:2009af,Abelev:2009jv,Chatrchyan:2012wg,Khachatryan:2010gv,CMS:2012qk,Abelev:2012ola,Aad:2012gla,phenix2015,star2015Eta,Adare:2015ctn}. In AA collisions, such long-range two-particle correlations have been associated with
the development of collective hydrodynamic flow, which transfers the azimuthal anisotropy in the initial energy density distribution to the final state momentum anisotropy
through strong rescatterings in the medium produced in such collisions~\cite{Ollitrault:1992bk,Andrade:2006yh,Alver:2010gr,Gale:2013da,Heinz:2013th}.
A recent study suggests that anisotropic escape probabilities may already produce large final-state anisotropies without the need for significant rescattering~\cite{He:2015hfa}.
Another possible mechanism proposed to account for the initial-state correlations is the color glass condensate (CGC), where the two-gluon density
is enhanced at small $\dphi$ over a wide $\deta$ range~\cite{Dumitru:2010iy,Dusling:2013oia}. However, to reproduce the magnitude of the ridge in
AA collisions, the CGC-based models also require a late-stage collective flow boost to produce the observed stronger angular collimation
effect~\cite{Gavin:2008ev,Dusling:2012iga}. As a purely initial-state effect,  the CGC correlations are expected to be independent of the formation of a
thermally equilibrated quark-gluon plasma, while the collective hydrodynamic flow requires a medium that is locally thermalized. The latter condition might
not be achieved in small systems.

Measurements at the LHC led to the discovery of a long-range ridge structure in small systems. The ridge has been observed in high-multiplicity
proton-proton (\pp)~\cite{PhysRevLett.116.172302,Khachatryan:2010gv,ppAtlas} and proton-lead (\pPb) collisions~\cite{CMS:2012qk,Abelev:2012ola,Aad:2012gla, Adam2016126}. A similar long-range structure was
also found in the most central deuteron-gold (dAu) and $^{3}$He-gold collisions at RHIC~\cite{phenix2015,star2015Eta,Adare:2015ctn}. To investigate
whether collective flow is responsible for the ridge in pPb collisions, multiparticle correlations were studied at the LHC~\cite{Chatrchyan:2013nka,atlasVn,cmsVn} in events with different multiplicities. The second harmonic anisotropy parameter, $v_2$, of the particle azimuthal distributions measured using four-, six-, eight-, or all-particle correlations were found to have the same value~\cite{cmsVn}, as expected in a system with global collective flow ~\cite{globalflow}.
In addition, the $v_2$ parameters of identified hadrons were measured as a function of transverse momentum (\pt) in  \pPb ~\cite{cmsPID,alicePID} and in dAu collisions~\cite{phenix2015}. The $v_2(\pt)$ distributions were found to be ordered by the particle mass, i.e., the distributions for the heavier particles are boosted to higher \pt, as expected from hydrodynamics, where the particles move with a common flow velocity.
The similarities between the correlations observed in the small systems and in heavy ion collisions suggest a common hydrodynamic origin~\cite{Bozek:2010pb,Bozek:2012gr,Chatrchyan:2013nka}. However it is still under investigation whether hydrodynamics can be applied reliably to \pp or \pA systems.

As predicted by hydrodynamics and CGC~\cite{Bozek:2014plb,Duraes:2015qoa}, as well as phenomenological models like EPOS~\cite{Pierog:2013ria}, the average transverse momentum, $\langle\pt\rangle$, of the produced particles should depend on
pseudorapidity. This pseudorapidity dependence of $\langle\pt\rangle$ could translate into a pseudorapidity dependence of the long-range correlations which also depend on $\pt$~\cite{Bozek:2015plb}.
While hydrodynamics predicts that the pseudorapidity dependence of $\langle\pt\rangle$ follows that of the charged particle
pseudorapidity density  $dN/d\eta$ which increases at negative pseudorapidity, in the CGC both a rising or a  falling trend
of  $\langle\pt\rangle$ with pseudorapidity may be possible~\cite{Duraes:2015qoa}. Thus, a measurement of the pseudorapidity dependence of the ridge may provide further insights into its origin.
The pseudorapidity dependence of the Fourier coefficients extracted using the long-range two-particle correlations could also be influenced by
event-by-event fluctuations of the initial energy density~\cite{PhysRevC.83.034911,PhysRevC.87.011901,PhysRevC.91.044904}. The pressure gradients that
drive the hydrodynamic expansion may differ in different pseudorapidity regions, causing a pseudorapidity-dependent phase shift in the event-plane orientation
determined from the direction of maximum particle emission. Evidence for such event-plane decorrelation has been found in pPb collisions~\cite{14012}.
Additional studies of the pseudorapidity dependence of the ridge may contribute to elucidating the longitudinal dynamics of the produced system.

The two-particle correlation measurement is performed using ``trigger'' and ``associated'' particles as described in Ref.~\cite{Chatrchyan:2011eka}. The trigger particles are defined as charged particles detected within a given $\pttrg$ range. The particle pairs are formed by associating each trigger particle with the remaining charged particles from a certain $\ptass$ range.
Typically, both particles are selected from a wide identical range of pseudorapidity, and therefore by construction the $\Delta\eta$ distribution is symmetric about $\deta = 0$~\cite{Chatrchyan:2013nka}.
Any $\Delta\eta$ dependence in the ridge correlation signal would be averaged out by the integration over the trigger and associated particle pseudorapidity distributions~\cite{Xu:2013sua}.
To gain further insights about the long-range ridge correlation in the \pPb system, in this paper we perform a $\deta$-dependent analysis by restricting the trigger particle to a narrow pseudorapidity range.
With this method, the combinatorial background resembles the single-particle density. Therefore, the correlation function in \pPb collisions is nonuniform in $\deta$.

The ridge correlation is often characterized by the Fourier coefficients $V_n$. The $V_n$ values are determined from a Fourier decomposition
of long-range two-particle \dphi\ correlation functions, given by:
\begin{linenomath}
\begin{equation}
\label{eq:Vn}
\frac{1}{N_\text{trig}}\frac{\rd N^\text{pair}}{\rd\Delta\phi} = \frac{N_\text{assoc}}{2\pi} \left[1+\sum\limits_{n} 2V_{n} \cos (n\Delta\phi)\right],
\end{equation}
\end{linenomath}
as described in Refs.~\cite{Chatrchyan:2011eka,Chatrchyan:2012wg}, where
$N^\text{pair}$ is the total number of correlated hadron pairs.
$N_\text{assoc}$ represents the total number of associated particles per trigger particle
for a given $(\pttrg, \ptass)$ bin.

 To remove short-range correlations from jets and other sources, a pseudorapidity separation may be applied between the trigger and associated particle; alternatively, the correlations in low multiplicity events may be measured and subtracted from those in high multiplicity events after appropriate scaling, to remove the short-range correlations, which are likely to have similar $\deta$-$\dphi$ shapes in high- and low-multiplicity collisions.
Both methods are used in this analysis.

The single-particle anisotropy parameters $v_n$ are extracted from the particle-pair Fourier coefficients $V_n$, assuming that they factorize~\cite{Aamodt:2011by}.
The $v_n$ values are then normalized by their lab frame mid-rapidity values
and are studied as a function of $\etacm$. These distributions are compared to
the normalized pseudorapidity distributions of the mean transverse momentum.

\section{CMS detector}
\label{sec:cmsdetector}

A detailed description of the CMS detector, together with a definition of the coordinate system used and the relevant kinematic variables, can be found in Ref.~\cite{Chatrchyan:2008zzk}.
The main results in this paper are based on data from the silicon tracker.
This detector consists of 1440 silicon pixel and 15\,148 silicon
strip detector modules, and is located in the 3.8\unit{T} magnetic field of the superconducting solenoid.
It measures the trajectories of the charged particles emitted within the pseudorapidity range $\abs{\etalab}< 2.5$,
and provides an impact parameter resolution of ${\sim}15\mum$ and a transverse
momentum resolution of about 1\% for particles with $\pt= 2\GeVc$, and  1.5\% for particles at $\pt= 100$\GeVc.

The electromagnetic calorimeter (ECAL) and the hadron calorimeter (HCAL) are also
located inside the solenoid. The ECAL consists of 75\,848
lead-tungstate crystals, arranged in a quasi-projective geometry and distributed in a
barrel region ($\abs{\etalab} < 1.48$) and two endcaps that extend up to $\abs{\etalab} = 3.0$.
The HCAL barrel and endcaps are sampling calorimeters composed of brass and
scintillator plates, covering $\abs{\etalab} < 3.0$. Iron/quartz fiber Cherenkov
Hadron Forward (HF) calorimeters cover the range $2.9 < \abs{\etalab} < 5.2$ on either
side of the interaction region.
The detailed MC simulation of the CMS detector response is based
on \GEANTfour~\cite{GEANT4}.

\section{Data samples and event selection}
\label{sec:data}

The data used are from \pPb collisions recorded by the CMS detector in 2013, corresponding to an integrated luminosity of  about 35\nbinv ~\cite{lumiRef}.
The beam energies were 4\TeV for protons and 1.58\TeV per nucleon
for lead nuclei, resulting in a center-of-mass energy per nucleon pair of $\rootsNN = 5.02\TeV$.
The direction of the higher-energy proton beam was initially set up to be clockwise, and then reversed.
Massless particles emitted at $\etacm = 0$ were detected at $\etalab = -0.465$ (clockwise proton beam) or at $\etalab = 0.465$ (counterclockwise proton beam) in the laboratory frame.
Both datasets were used in this paper. The data in which the proton beam traveled clockwise were reflected about $\etalab = 0$ and combined with the rest of the  data, so that the proton beam direction is always associated with the positive $\etalab$ direction.

The online triggering, and the offline reconstruction and selection follow the same procedure as described in Ref.~\cite{Chatrchyan:2013nka}.
Minimum-bias events were selected by requiring that at least one track with $\pt>0.4\GeVc$ was found in the pixel tracker for a \pPb bunch crossing.
Because of hardware limits on the data acquisition rate, only a small fraction ($10^{-3}$) of all minimum bias triggered events were recorded (i.e., the trigger was ``prescaled'').
 The high-multiplicity triggers were implemented using the Level-1 (L1) trigger and High Level Trigger (HLT) to enhance high multiplicity events that are of interest for the particle correlation studies.
At L1, two event streams were triggered by requiring the total transverse energy summed over ECAL and HCAL to be greater than 20 or 40\GeVc.
Charged tracks were then reconstructed online at the HLT using the three layers of pixel detectors, and requiring a track origin within a cylindrical region of 30 cm length along the beam and 0.2\unit{cm} radius perpendicular to the beam~\cite{Stenson:2010xx}.

In the offline analysis, hadronic collisions were selected by requiring at least 3\GeVc of total energy in at least one HF
calorimeter tower on each side of the interaction region (positive and negative $\etalab$).
Events were also required to contain at least one reconstructed primary
vertex within 15\unit{cm} of the nominal interaction point along the beam axis ($z_\text{vtx}$)
and within 0.15\unit{cm} distance transverse to the beam trajectory.

The \pPb instantaneous luminosity provided by the LHC in the 2013 \pPb run resulted in approximately a 3\% probability that at least one additional interaction occurs in the same bunch crossing, i.e. pileup events. A pileup rejection procedure~\cite{Chatrchyan:2013nka} was applied to select clean, single-vertex \pPb events. The residual fraction of pileup events was estimated to be no more than 0.2\% for the highest multiplicity \pPb interactions studied in this paper~\cite{Chatrchyan:2013nka}.
Based on simulations using the \HIJING~\cite{Gyulassy:1994ew}
and the \textsc{epos}~\cite{Porteboeuf:2010um} event generators, these event selections
have an acceptance of 94--97\% for \pPb interactions
that have at least one primary particle with $E>3$\GeV in both $\etalab$ ranges of
$-5<\etalab<-3$ and $3<\etalab<5$.
The charged-particle information was recorded in the silicon tracker and the tracks were reconstructed within the pseudorapidity range $\abs{\etalab}< 2.5$.

A reconstructed track was considered as a primary track candidate if the impact parameter
significance $d_{xy}/\sigma(d_{xy})$ and the significance of $z$ separation between the
track and the best reconstructed primary vertex (the one associated with the largest
number of tracks, or best $\chi^{2}$ probability if the same number of tracks was found)
$d_z/\sigma(d_z)$ are both less than 3.
 In order to remove tracks with poor momentum estimates,
the relative uncertainty in the momentum measurement $\sigma(\pt)/\pt$ was required
to be less than 10\%.
To ensure high tracking efficiency and to reduce the rate of misreconstructed tracks, primary tracks with $\abs{\etalab}<2.4$ and $\pt>0.3\GeVc$ were used in the analysis.

The events are classified by \noff, the measured number of primary tracks within $\abs{\etalab}<2.4$ and $\pt>0.4\GeVc$ (a $\pt$ cutoff of 0.4\GeVc was used in the multiplicity determination to match the HLT requirement), in a method similar to the approach used in Refs.~\cite{CMS:2012qk,Khachatryan:2010gv}.
The high- and low-multiplicity events in this paper are defined by $220\leq\noff<260$ and $2\leq\noff<20$, respectively.
The high-multiplicity selection corresponds to an event fraction of $3.4\times 10^{-6}$ of the events.
 Data from the minimum bias trigger are used for low-multiplicity event selection, while the high-multiplicity triggers with online multiplicity thresholds of 100, 130, 160, and 190 are used for high multiplicity events~\cite{Chatrchyan:2013nka}.

\section{Analysis procedure}\label{sec:}

The dihadron correlation is quantified by azimuthal angle $\phi$ and pseudorapidity differences between the two particles.
\begin{linenomath}
\begin{equation*}
\label{detadphi}
\Delta\phi=\phi_\text{assoc}-\phi_\text{trig},\quad \Delta\eta=\etaassoc-\etatrg,
\end{equation*}
\end{linenomath}
where $\phi_\text{assoc}$ and $\etaassoc$ are the associated particle coordinates and $\phi_\text{trig}$ and $\etatrg$ are the trigger particle coordinates, both measured in the laboratory frame.
The per-trigger normalized associated particle yield is defined by:
\begin{linenomath}
\begin{equation*}
\label{eq:signal}
S(\Delta\eta,\ \Delta\phi) = \frac{1}{N_\text{trig}}\frac{\rd^{2}N}{\rd\Delta\eta\, \rd\Delta\phi}.
\end{equation*}
\end{linenomath}

Unlike in previous studies ~\cite{Adams:2005ph,Alver:2009id,Abelev:2009af,Khachatryan:2010gv,CMS:2012qk,Abelev:2012ola,Aad:2012gla}, the trigger particles in this analysis are restricted to two narrow $\etalab$ windows: $-2.4<\etatrg<-2.0$ (Pb-side) and $2.0<\etatrg<2.4$ (p-side). The associated particles are from the entire measured $\etalab$ range of $-2.4<\etaassoc<2.4$.

The associated particles are weighted by the inverse of the
efficiency factor, $\varepsilon_\text{trk}(\etalab,\pt)$,
as a function of the track's pseudorapidity and $\pt$~\cite{Chatrchyan:2011eka}.
The efficiency factor accounts for the detector
acceptance $A(\etalab,\pt)$, the reconstruction efficiency $E(\etalab,\pt)$,
and the fraction of misidentified tracks, $F(\etalab,\pt)$,
\begin{linenomath}
\begin{equation*}
\varepsilon_\text{trk}(\etalab,\pt) = \frac{A E}{1-F}\,.
\end{equation*}
\end{linenomath}

The corresponding correction function is obtained from a {\PYTHIA6 }~(tune Z2)~\cite{PYTHIA} plus \GEANTfour \cite{GEANT4} simulation.
\subsection{Quantifying the jet contributions}

\begin{figure*}[thb]
\centering
\includegraphics[width=0.48\textwidth]{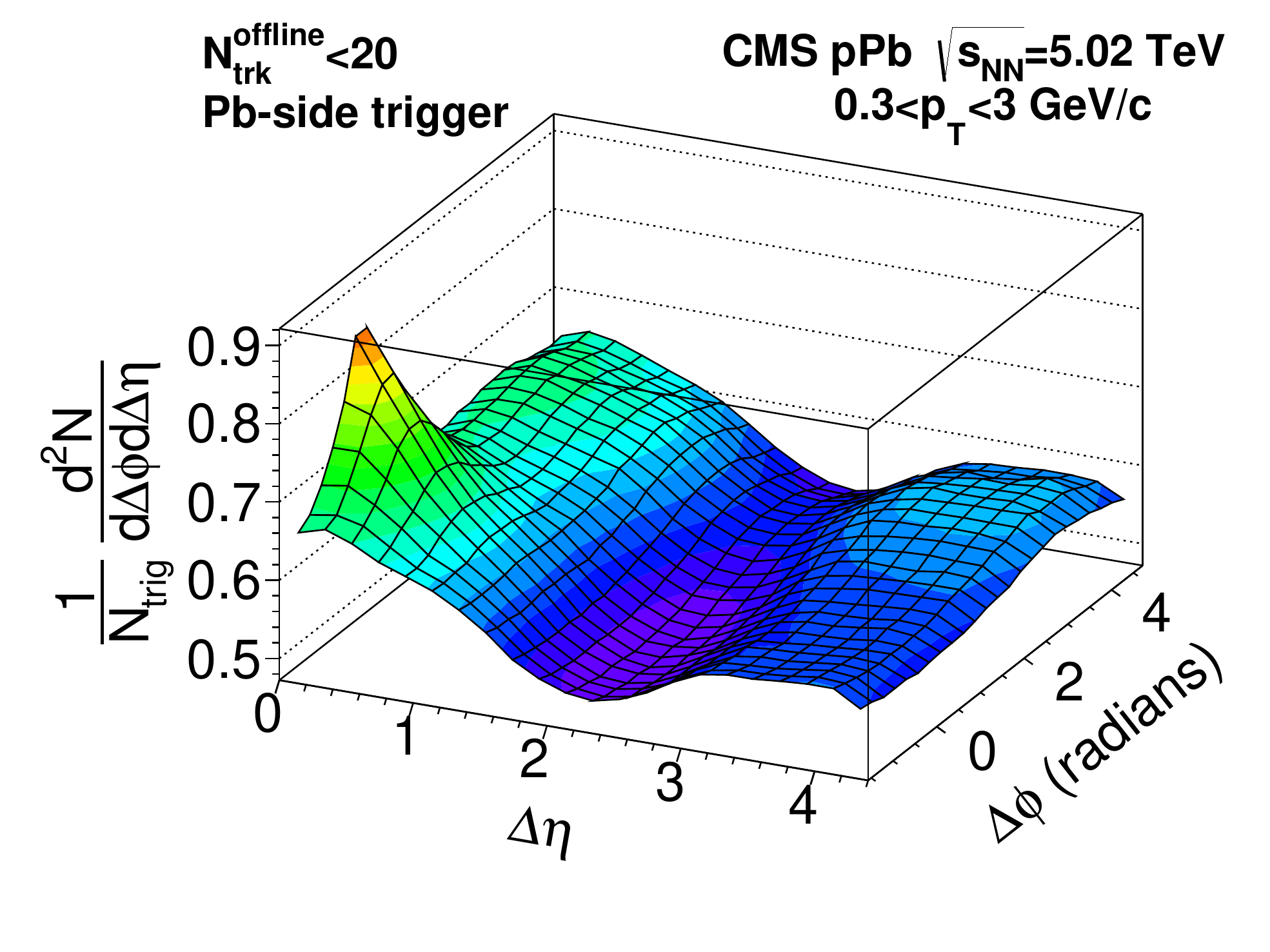}
\includegraphics[width=0.48\textwidth]{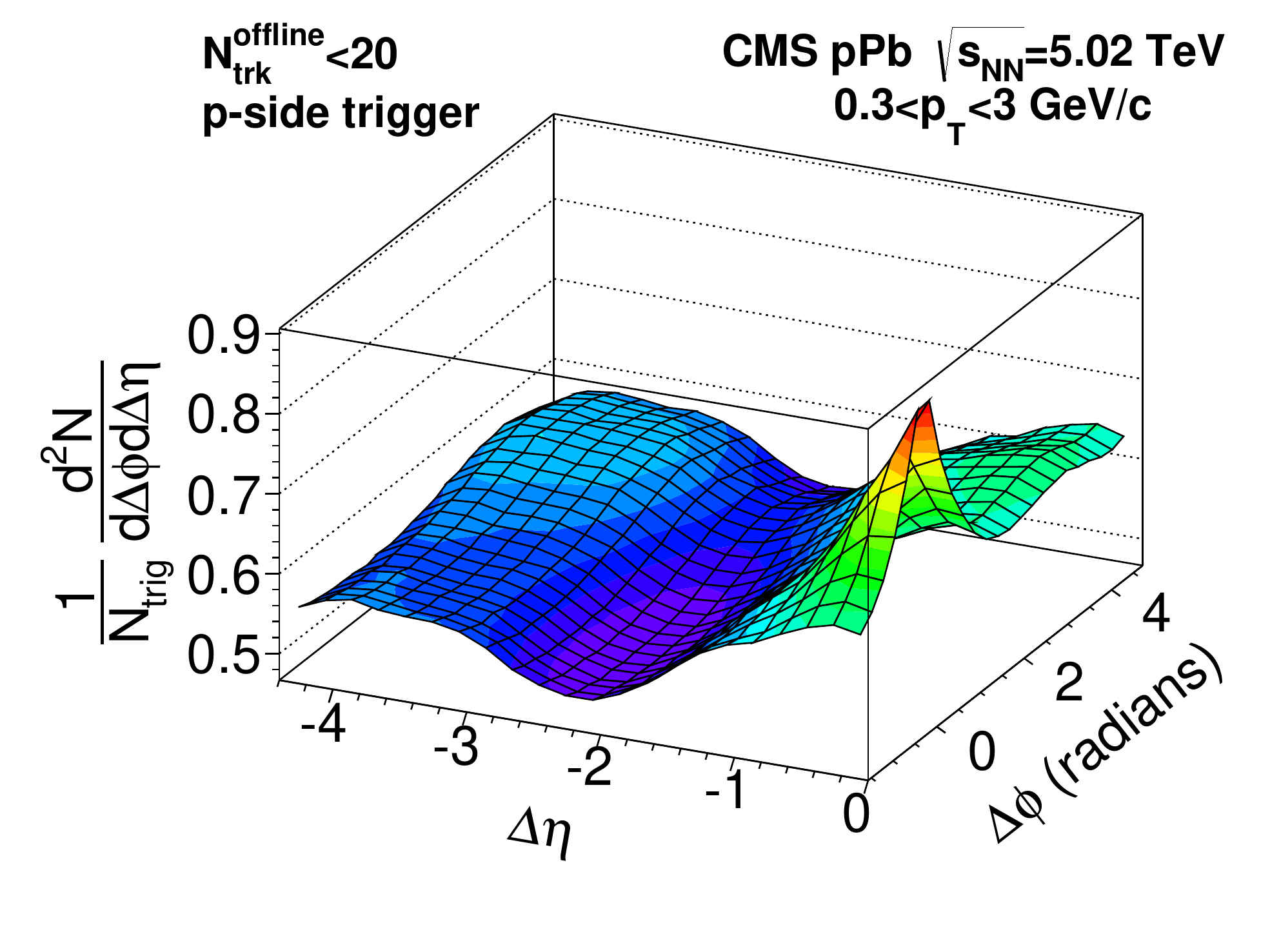}
\includegraphics[width=0.48\textwidth]{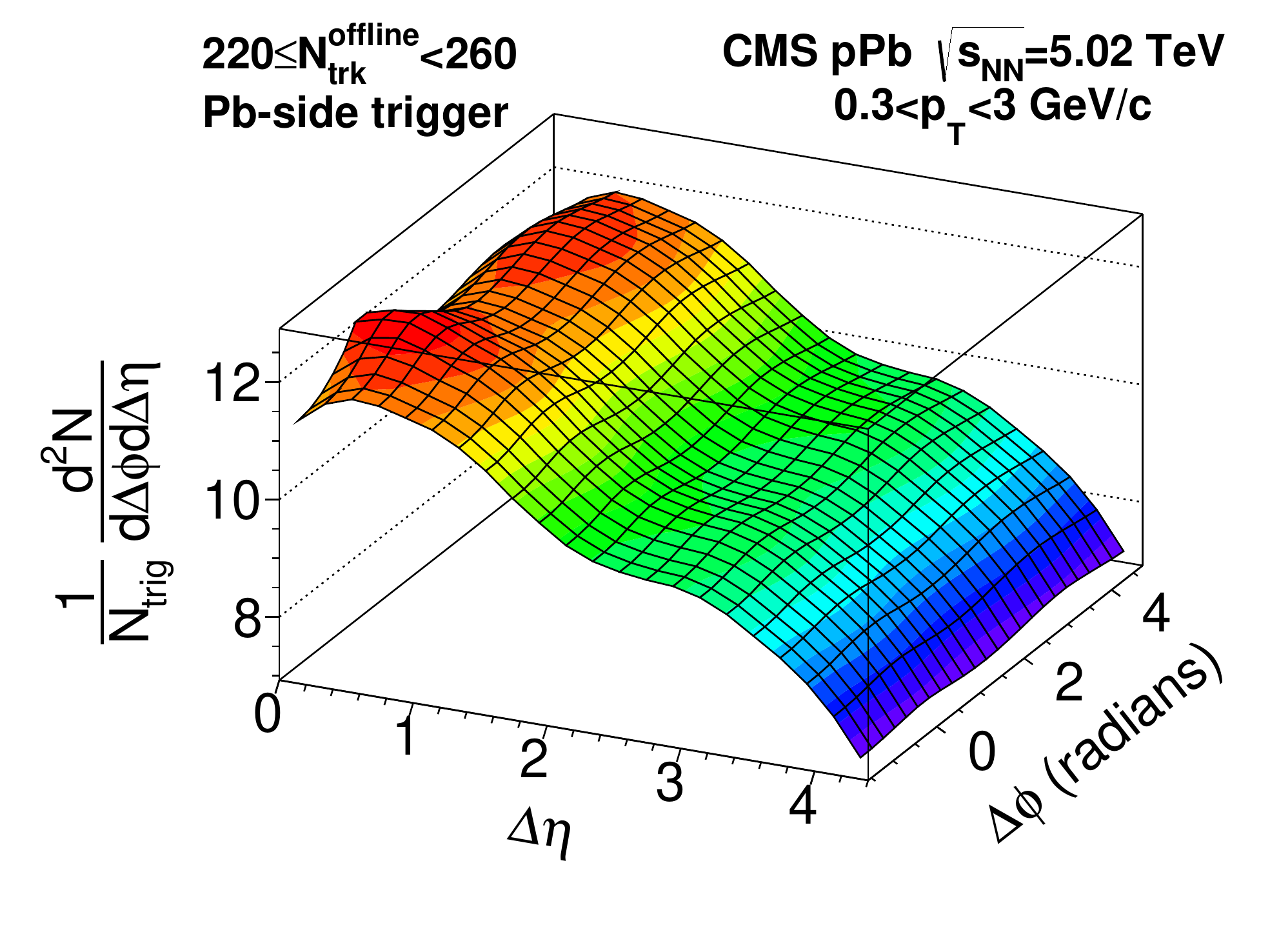}
\includegraphics[width=0.48\textwidth]{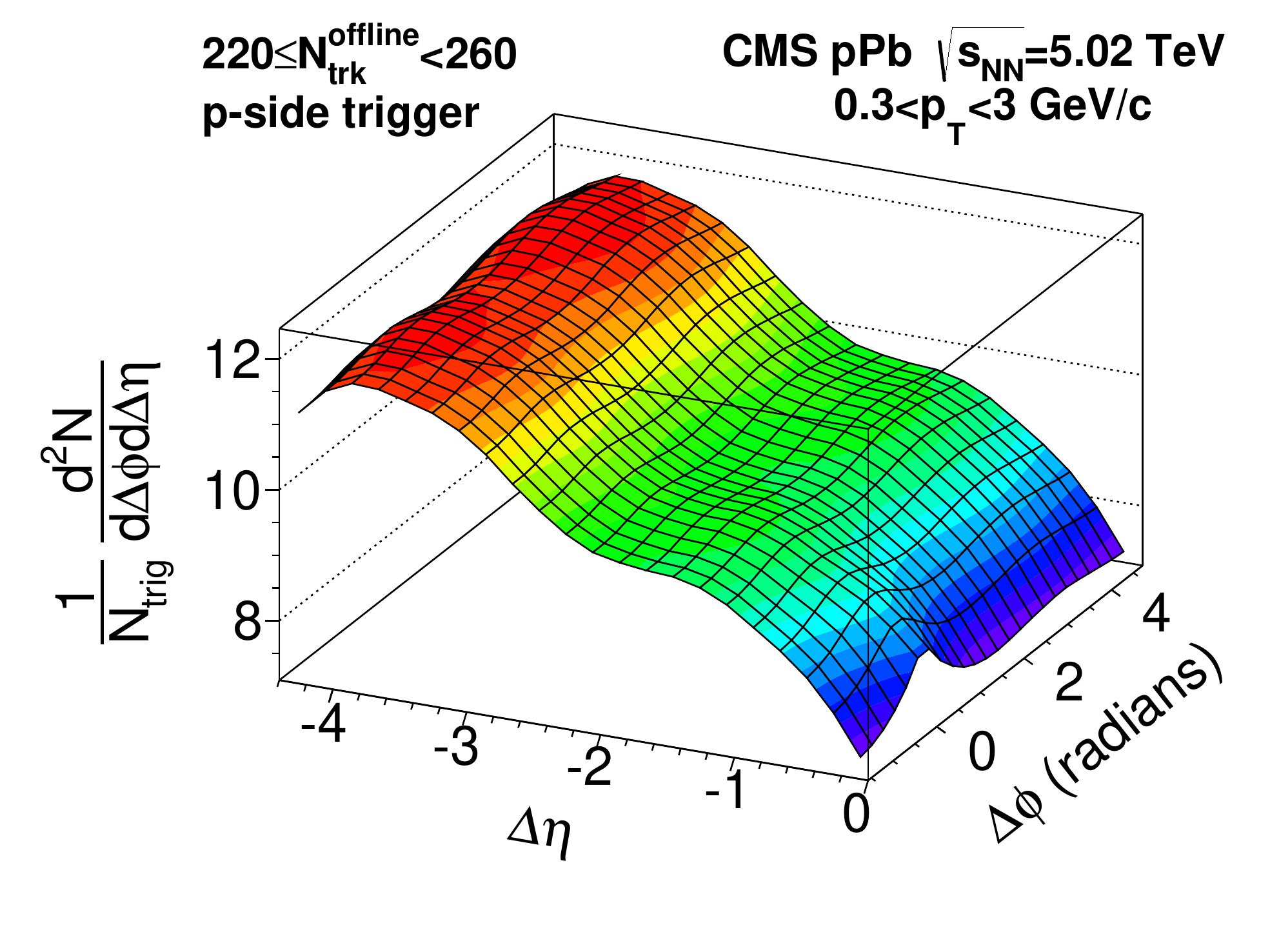}
\caption{ \label{fig:pPb_corrYield_2D}(Color online) Efficiency-corrected 2D associated yields with Pb-side trigger particle ($-2.4<\etatrg<-2.0$, left panels) and p-side trigger particle ($2.0<\etatrg<2.4$, right panels) in low-multiplicity ($2\leq\noff<20$, upper panels) and high-multiplicity ($220\leq\noff<260$, lower panels) are shown for \pPb collisions at $\rootsNN = 5.02\TeV$. The associated and trigger particle $\pt$ ranges are both $0.3<\pt<3\GeVc$.}
\end{figure*}

Figure~\ref{fig:pPb_corrYield_2D} shows the two-dimensional (2D) correlated yield for the two trigger particle pseudorapidity windows in low and high multiplicity events. The same $\pt$ range of $0.3<\pt<3.0\GeVc$ is used for trigger and associated particles.
The peak at $(0,0)$ is the near-side jet-like structure.
In the high multiplicity events, one can notice a ridge-like structure in $\abs{\deta}$ at $\dphi=0$ atop the high combinatorial background. A similar extensive structure can also be seen on the away side $\dphi=\pi$, which contains the away-side jet.
Unlike correlation functions from previous studies, the correlated yield is asymmetric in $\Delta\eta$; it reflects the asymmetric single particle $\rd N/\rd\eta$ distribution in the \pPb system.

\begin{figure*}[thb]
\centering
	\includegraphics[width=\textwidth]{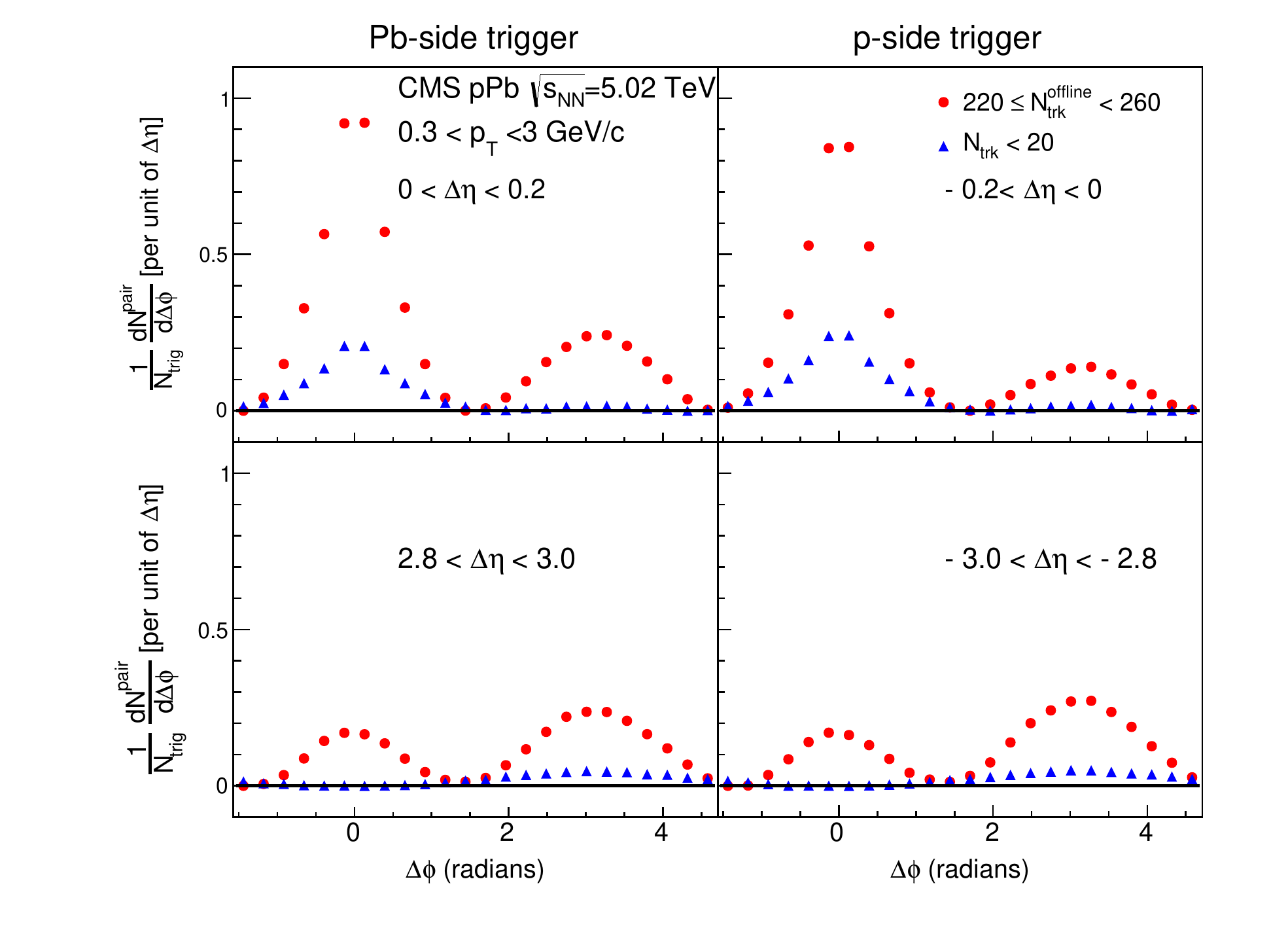}
\caption{\label{fig:dphi} (Color online) Examples of the distribution of the associated yields after ZYAM subtraction for both low-multiplicity ($2\leq\noff<20$, blue triangles) and high-multiplicity ($220\leq\noff<260$, red circles) are shown for \pPb collisions at $\rootsNN = 5.02\TeV$. The results for Pb-side (left panels) and p-side (right panels) trigger particles are both shown; small $\deta$ in the upper panels and large $\abs{\deta}$ in the lower panels. The trigger and associated particle \pt ranges are both $0.3<\pt<3\GeVc$.}
\end{figure*}

The $\dphi$ distribution of the associated yield is projected within each $\Delta\eta$ bin (with a bin width of 0.2).
Before quantifying jet contributions, the zero-yield-at-minimum (ZYAM) technique~\cite{Ajitanand:2005jj} is used to subtract a uniform background in $\Delta\phi$.
To obtain the ZYAM background normalization, the associated yield distribution is first projected into the range of $0<\Delta\phi<\pi$, and then scanned to find the minimum yield within a $\dphi$ window of $\pi/12$ radians. This minimum yield is treated as the ZYAM background. The ZYAM background shape as a function of $\deta$ is similar to the shape of the single particle density.

After ZYAM subtraction, the signal will be zero at the minimum.
For example, the $\dphi$ distributions in high- and low-multiplicity collisions are depicted in Fig.~\ref{fig:dphi} for two, short-($0<\abs{\deta}<0.2$) and long-range($2.8<\abs{\deta}<3.0$), $\deta$ bins.
They are composed of two characteristic peaks:
one at $\Delta\phi=0$ (near-side) and the other at $\Delta\phi=\pi$ (away-side), with a minimum valley between the two peaks.
For low-multiplicity collisions at large $\deta$, no near-side peak is observed.

\begin{figure*}[thb]
\centering
\includegraphics[width=\textwidth]{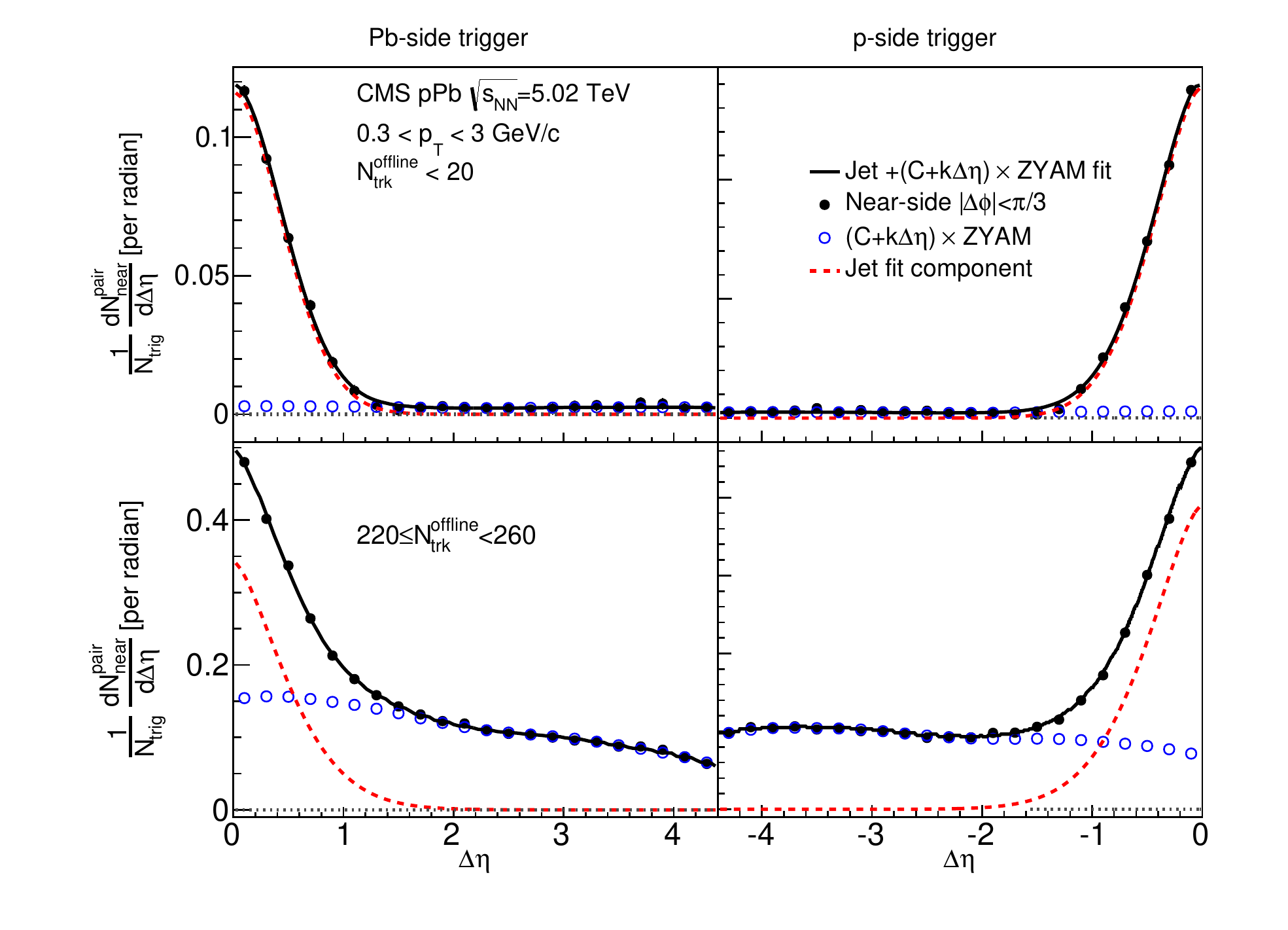}
\caption{\label{fig:near_fit_pPb}(Color online) The near-side ($|\dphi|<\pi/3$) correlated yield after ZYAM subtraction in low-multiplicity $2\leq\noff<20$ (upper panels) and high-multiplicity $220\leq\noff<260$ (lower panels) are shown for \pPb collisions at $\rootsNN = 5.02\TeV$. The trigger and associated particle $\pt$ ranges are both $0.3<\pt<3\GeVc$. The trigger particles are restricted to the Pb-side ($-2.4<\etatrg<-2.0$, left panels) and the p-side ($2.0<\etatrg<2.4$, right panels), respectively. Fit results using Eq.~(\ref{eq:nearfit}) (black solid curves) are superimposed; the red dashed curve and the blue open points are the two fit components, jet and ridge, respectively.
}
\end{figure*}

First, the $\deta$ dependence of the correlated yield is analyzed. In each $\deta$ bin, the correlated yield is averaged within the near side ($|\Delta\phi|<\pi/3$. The correlated yield reaches a minimum at around $\pi/3$). The near-side averaged correlated yield per radian, $(1/N_\text{trig}) (\rd N)/(\rd\deta)$, is shown as a function of $\deta$ in Fig.~\ref{fig:near_fit_pPb}.
In low-multiplicity collisions, the near-side $\deta$ correlated yield is consistent with zero at large $\deta$. This indicates that the near side in low-multiplicity \pPb collisions is composed of only a jet component after ZYAM subtraction. In high-multiplicity collisions, an excess of the near-side correlated yield is seen at large $\deta$ and it is due to the previously observed ridge~\cite{CMS:2012qk}.

In order to quantify the near-side jet contribution, the near-side correlation function is fitted with a two-component functional form:
\begin{linenomath}
\ifthenelse{\boolean{cms@external}}{
\begin{multline}
\label{eq:nearfit}
\frac{1}{N_\text{trig}}\,\frac{\rd N_\text{near}(\deta)}{\rd\deta}=\frac{Y\beta}{\sqrt{2}\sigma\Gamma(1/2\beta)}\exp\left[-\left(\frac{\Delta\eta^{2}}{2\sigma^{2}}\right)^{\beta}\right]\\
+(C+k\Delta\eta)\mathrm{ ZYAM}(\Delta\eta).
\end{multline}
}{
\begin{equation}
\label{eq:nearfit}
\frac{1}{N_\text{trig}}\,\frac{\rd N_\text{near}(\deta)}{\rd\deta}=\frac{Y\beta}{\sqrt{2}\sigma\Gamma(1/2\beta)}\exp\left[-\left(\frac{\Delta\eta^{2}}{2\sigma^{2}}\right)^{\beta}\right]+(C+k\Delta\eta)\mathrm{ ZYAM}(\Delta\eta).
\end{equation}
}
\end{linenomath}
The first term represents the near-side jet; $Y$ is the correlated yield, and $\sigma$ and $\beta$ describe the correlation shape. Neither a simple Gaussian nor an exponential function describes the jet-like peak adequately. However, a generalized Gaussian form as in Eq.~(\ref{eq:nearfit}) is found to describe the data well.
The second term on the right-hand side of Eq.~(\ref{eq:nearfit}) represents the ridge structure. Since the ridge is wide in $\Delta\eta$ and may be related to the bulk medium, its shape is modeled as dominated by the underlying event magnitude, $\mathrm{ZYAM}(\deta)$. However, the background shape multiplied by a constant is not adequate to describe the ridge in high multiplicity events. Instead, the background shape multiplied by a linear function in $\Delta\eta$, as in Eq.~(\ref{eq:nearfit}), can fit the data well, with reasonable $\chi^2/\mathrm{ndf}$ (where ndf is the number of degree of freedom) (see Table~\ref{tab:fit_para}). Here $C$ quantifies the overall strength of the ridge yield relative to the underlying event, and $k$ indicates the $\deta$ dependence of the ridge in addition to that of the underlying event.

\begin{table}
\topcaption{\label{tab:fit_para} Summary of fit parameters for low- and high-\noff\ ranges in \pPb collisions. }
\centering
\begin{scotch}{  l  x  x }
\multicolumn{3}{c}{\rule[-0.4\ruleht]{0pt}{1.25\ruleht}$\noff <20$} \\
\hline
 Parameter                          & \multicolumn{1}{c}{Pb-side trigger} & \multicolumn{1}{c}{p-side trigger}\\[1.2ex]
 Y                                  &0.130,0.003      & 0.156,0.003\\
 $\sigma$                           &0.445,0.011       &0.446,0.010\\
 $\beta$                            &0.943,0.057      & 0.870,0.043\\
 $C$                                  & 0.0045,0.0009    & 0.0045,0.0010\\
 $k$                                  & \multicolumn{1}{l}{0 (Fixed)}  & \multicolumn{1}{l}{0 (Fixed)}\\
 $\chi^2/\mathrm{ndf}$                           & \multicolumn{1}{l}{0.279}   & \multicolumn{1}{l}{0.459}\\
\hline
\multicolumn{3}{c}{\rule[-0.4\ruleht]{0pt}{1.25\ruleht}$220 \leq \noff <260$}\\
\hline
 Parameter                          & \multicolumn{1}{c}{Pb-side trigger} & \multicolumn{1}{c}{p-side trigger}\\[1.2ex]
 Y                                    & 0.401,0.011&0.489,0.011\\
 $\sigma$                             &0.457,0.008 &0.492,0.007 \\
 $\beta$                              & 0.757,0.003 &0.782,0.025 \\
 $C$                                    & 0.0137,0.0004 & 0.0098,0.0004 \\
 $k$                                    & - 0.0011,0.0001& 0.0002,0.0001 \\
 $\chi^2/\mathrm{ndf}$                             & \multicolumn{1}{l}{1.074}   & \multicolumn{1}{l}{0.463}\\
\end{scotch}
\end{table}

The fits using Eq.~(\ref{eq:nearfit}) are superimposed in Fig.~\ref{fig:near_fit_pPb} and the fit parameters are shown in Table~\ref{tab:fit_para}. For low-multiplicity collisions, the $k$ parameter is consistent with zero and, in the fit shown, it is set to zero.
For high-multiplicity collisions, the $C$ parameter is positive, reflecting the finite ridge correlation, and the $k$ parameter is nonzero, indicating that the ridge does not have the same $\deta$ shape as the underlying event.
As already shown in Fig.~\ref{fig:near_fit_pPb}, the ridge (correlated yield at large $\deta$) is not constant but $\deta$-dependent.

The fitted $Y$ parameter shows that the jet-like correlated yield in high-multiplicity collisions ($Y_{ 220 \leq \noff <260}$) is larger than that in low-multiplicity collisions ($Y_{ \noff <20}$). The ratio is
\begin{linenomath}
\ifthenelse{\boolean{cms@external}}
{\begin{multline}
 \alpha=Y_{ 220 \leq \noff <260}/Y_{ \noff <20} \\=
 \begin{cases}
    3.08\pm 0.11^{+0.96}_{-0.31}       & \quad \text{for Pb-side triggers};           \\
    3.13\pm 0.09^{+0.28}_{-0.28}  & \quad \text{for p-side triggers},            \\
  \end{cases}
\label{eq:alphaEq222}
\end{multline}
}
{\begin{equation}
 \alpha=Y_{ 220 \leq \noff <260}/Y_{ \noff <20} =
 \begin{cases}
    3.08\pm 0.11^{+0.96}_{-0.31}       & \quad \text{for Pb-side triggers};           \\
    3.13\pm 0.09^{+0.28}_{-0.28}  & \quad \text{for p-side triggers},            \\
  \end{cases}
\label{eq:alphaEq222}
\end{equation}
}
\end{linenomath}
 where the $\pm$ sign is followed by the statistical uncertainty from the fit. The upper ``$+$'' and lower ``$-$'' are followed by the systematic uncertainty, which is obtained by fitting different functional forms, such as Gaussian and exponential functions, and by varying the $\deta$ range to calculate the ZYAM value.

 The $\alpha$ values are used as a scaling factor when correlations from low-multiplicity collisions are removed in determining the Fourier coefficients in high-multiplicity events.

\subsection{Fourier coefficients}
For each $\deta$ bin,
the azimuthal anisotropy harmonics, $V_n$, can be calculated from the two-particle correlation $\dphi$ distribution,
\begin{linenomath}
\begin{equation*}
V_n=\langle\cos n \Delta\phi\rangle.
\label{eq:2pVn}
\end{equation*}
\end{linenomath}

The $\langle\rangle$ denotes the averaging over all particles and all events.
At large $\deta$, the near-side jet contribution is negligible, but the away-side jet still contributes. The jet contributions may be significantly reduced or eliminated by subtracting the low-multiplicity collision data, via a prescription described in Ref.~\cite{Chatrchyan:2013nka},
\begin{linenomath}
\begin{equation}
\label{eq:VnSub}
V_n^\text{sub} = V_n^\mathrm{HM} - V_n^\mathrm{LM} \frac{N_\text{assoc}^\mathrm{LM}}{N_\text{assoc}^\mathrm{HM}}\alpha.
\end{equation}
\end{linenomath}
Here LM and HM stand for low-multiplicity and high-multiplicity, respectively. $N_\text{assoc}^\mathrm{HM}$ and $N_\text{assoc}^\mathrm{LM}$ are the associated particle multiplicities in a given pseudorapidity bin, and $V_n^\mathrm{HM}$ and $V_n^\mathrm{LM}$ are the Fourier coefficients in high- and low-multiplicity collisions, respectively. The $\alpha$ value is obtained from Eq.~(\ref{eq:alphaEq222}).
This procedure to extract $V_{n}$ is tested by studying the \pPb collisions generated by the \HIJING 1.383 model ~\cite{Gyulassy:1994ew}.
The basic \HIJING model has no flow, so a flow-like signal is added~\cite{afterburnerflow} by superimposing an azimuthal modulation on the distributions of the produced particles.
The measured $V_2$ using Eq.~(\ref{eq:VnSub}) is consistent with the input flow value within a relative 5\% difference.

To quantify the anisotropy dependence as a function of $\etalab$, assuming factorization, \breakhere$V_n(\etatrg, \etaassoc) = v_n(\etatrg)v_n (\etaassoc)$, a self-normalized anisotropy is calculated from the Fourier coefficient $V_{n}$.
\begin{linenomath}
\begin{equation}
\label{eq:vnNormalize}
\frac{v_n(\etaassoc)}{v_n(\etaassoc=0)}=\frac{V_n(\etaassoc)}{V_n(\etaassoc=0)}.
\end{equation}
\end{linenomath}
Here the $\etaassoc$ is directly calculated from $\deta$, assuming the trigger particle is at a fixed $\etalab$ direction
\begin{linenomath}
\begin{equation}
\label{eq:etaShift}
\etaassoc=\deta+\etatrg,
\end{equation}
\end{linenomath}
in which $\etatrg=-2.2\,(2.2)$ for the Pb-side (p-side) trigger. Hereafter, we write only $\etalab$, eliminating the superscript `assoc' from $\etaassoc$.

To avoid short-range correlations that remain even after the subtraction of the low-multiplicity events, only correlations with large $\abs{\deta}$ are selected to construct the $v_n$ pseudorapidity distributions.
\section{Systematic uncertainties}
\label{sec:systematic}

The systematic uncertainties in the Fourier coefficient $V_n$ are estimated from the following sources:
the track quality requirements by comparing loose and tight selections; bias in the event selection from the HLT trigger, by using different high-multiplicity event selection criteria; pileup effect, by requiring a single vertex per event; and the event vertex position, by selecting events from different z-vertex ranges.
In the low multiplicity $V_n$ subtraction, the jet ratio parameter $\alpha$ is applied.
The systematic uncertainties in $\alpha$ are assessed by using fit functions different from Eq.~(\ref{eq:nearfit}), as well as by varying the $\deta$ range when obtaining the ZYAM value. This systematic effect is included in the final uncertainties for the multiplicity-subtracted $V_n$.
In addition, the effect of reversing the beam direction is studied. This is subject to the same systematic uncertainties already described above; thus it is not counted in the total systematic uncertainties, but is used as a cross-check.

The estimated uncertainties from the above sources are shown in Table~\ref{tab:syst-table-new}. Combined together, they give a total uncertainty of 3.9\% and 10\% for $V_2$ and $V_3$ coefficients, respectively, as determined without the subtraction of signals from low-multiplicity events. For low-multiplicity-subtracted results, the systematic uncertainties rise to 5.8\% and 15\%, respectively.

The systematic uncertainties from the track-quality and jet-ratio selection are correlated among the pseudorapidity bins,
so they cancel in the self-normalized anisotropy parameter, \breakhere$v_n(\etalab)/v_n(\etalab=\nobreak 0)$.
The systematic uncertainties in other sources are treated as completely independent of pseudorapidity and are propagated in $v_n(\etalab)/v_n(\etalab\nobreak =0)$.
The estimated systematic uncertainties in $v_2(\etalab)/v_2(\etalab=0)$ and $v_3(\etalab)/v_3(\etalab=0)$ without low multiplicity subtraction are estimated to be 3.6\% and 10\%, respectively. For low-multiplicity-subtracted results, the systematic uncertainties rise to 5.7\% and 14\%, respectively  .

\begin{table*}[htb]
\topcaption{\label{tab:syst-table-new} Summary of systematic uncertainties in the second and third Fourier harmonics in \pPb collisions. The label ``low-mult sub'' indicates the low-multiplicity subtracted results, while "no sub" indicates the results without subtraction.}
\centering
 \begin{scotch}{ccccc}
\multicolumn{5}{c}{\rule[-0.4\ruleht]{0pt}{1.25\ruleht} $220\leq\noff <  260$ } \\
\hline
 Source                                         & $V_2$ (no sub)   &   $V_2$ (low-mult sub)   & $V_3$ (no sub)   &   $V_3$ (low-mult sub)\\[1.2ex]
 Track quality requirement                         & 3.0\%  & 3.0\% & 7.0\% & 11.0\%\\
 HLT trigger bias                                  & 2.0\%  & 2.5\% & 2.0\% & 2.5\% \\
 Effect from pileups                               & 1.5\%  & 3.0\% & 3.5\% & 3.5\% \\
 Vertex dependence                                 & 0.5\%  & 1.0\% & 6.0\% & 9.0\% \\
 Jet ratio                                         &\NA     & 3.0\% &\NA     & 3.0\% \\[1.2ex]
 Total                                             & 3.9\%  & 5.8\% & 10\%  & 15\% \\
\end{scotch}
\end{table*}

\section{Results}

\begin{figure*}[thb]
\centering
\includegraphics[width=\textwidth]{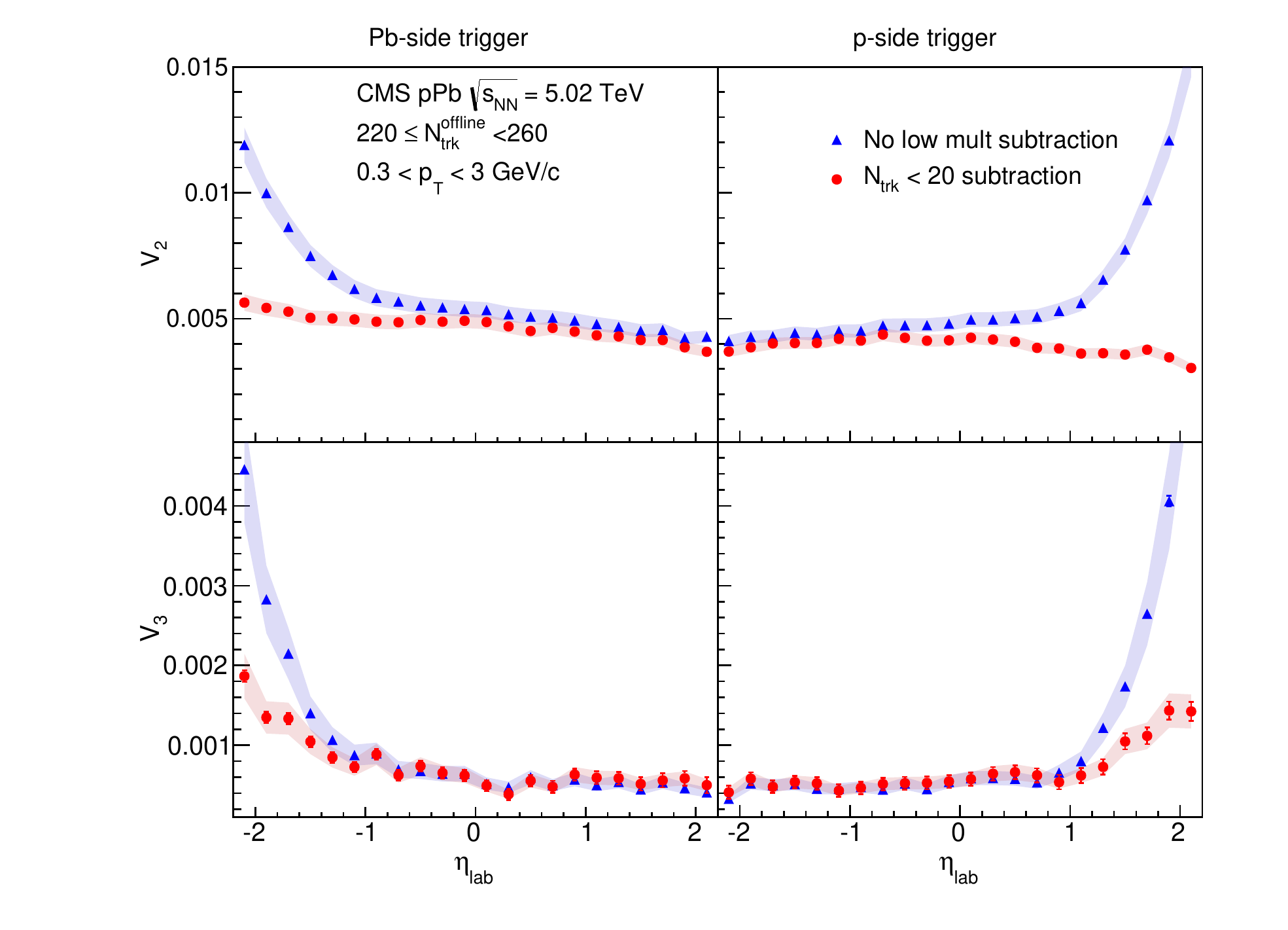}
\caption{(Color online) Fourier coefficients, $V_2$ (upper) and $V_3$ (lower), of two-particle azimuthal correlations in high-multiplicity collisions ($220\leq\noff<260$) with (circles) and without (triangles) subtraction of low-multiplicity data, as a function of $\etalab$. Left panel shows data for Pb-side trigger particles and the right panel for the p-side. Statistical uncertainties are mostly smaller than point size; systematic uncertainties are $3.9\%$ and $10\%$ for $V_2$ and $V_3$ without low-multiplicity subtraction, $5.8\%$ and $15\%$ for $V_2$ and $V_3$ with low-multiplicity subtraction, respectively. The systematic uncertainties are shown by the shaded bands.}
\label{fig:V2_pPb}
\end{figure*}

The $V_2$ and $V_3$ values in high-multiplicity collisions for Pb-side and p-side trigger particles are shown in Fig.~\ref{fig:V2_pPb}. The strong peak is caused by near-side short-range jet contributions.
The Fourier coefficients, $V_2^\text{sub}$ and $V_3^\text{sub}$, after the low-multiplicity data are subtracted, are also shown. The short-range jet-like peak is largely reduced, but may not be completely eliminated due to different near-side jet-correlation shapes for high- and low-multiplicity collisions. The long-range results are not affected by the near-side jet, but the away-side jet may still contribute if its shape is different in high- and low-multiplicity collisions or if its magnitude does not scale according to $\alpha$.

By self-normalization via Eq.~(\ref{eq:vnNormalize}), the Fourier coefficient from both trigger sides can be merged into a single distribution by combining the negative and positive $\etalab$ range. The lab frame central value $\etalab=0$ is used so that the separation of the central value to both $\etatrg$ is the same. In this way, possible contamination from jets is kept at the same level as a function of $\etalab$. This is more important for the Fourier coefficients determined without the subtraction of the low-multiplicity data.

\begin{figure}[thb]
\centering
\includegraphics[width=0.48\textwidth]{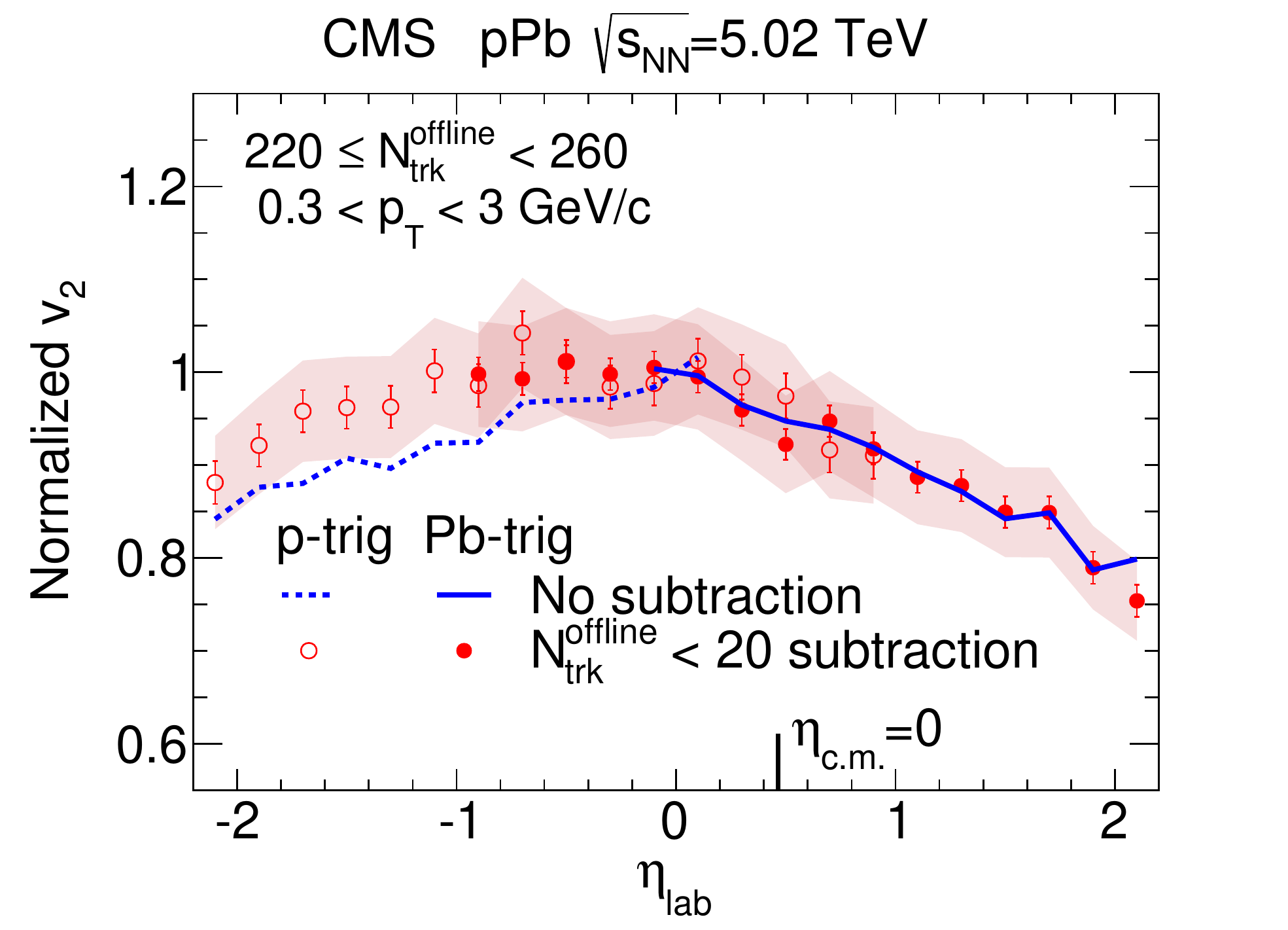}
\includegraphics[width=0.48\textwidth]{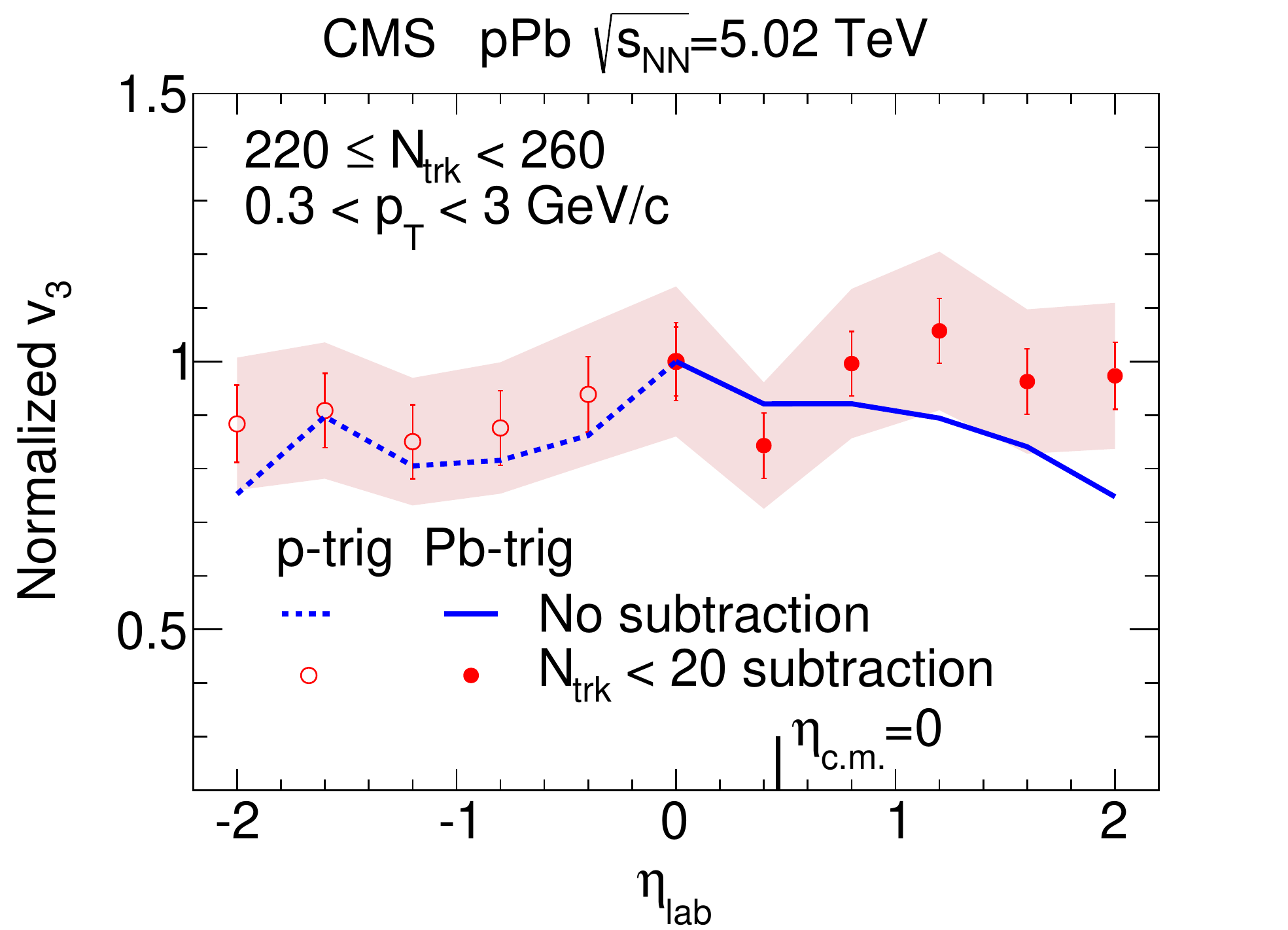}
\caption{(Color online) Self-normalized anisotropy parameters, $v_2(\etalab)/v_2(\etalab=0)$ (\cmsLeft panel) and $v_3(\etalab)/v_3(\etalab=0)$ (\cmsRight panel), as a function of $\etalab$. Data points (curves) are results with (without) low-multiplicity data subtraction; filled circles and solid lines are from the Pb-side trigger. Open circles and dashed lines are from the p-side trigger. The bands show systematic uncertainties of $\pm5.7\%$ and $\pm14\%$ for $v_2(\etalab)/v_2(\etalab=0)$ and $v_3(\etalab)/v_3(\etalab=0)$, respectively. The systematic uncertainties in $v_n(\etalab)/v_n(\etalab=0)$ without subtraction are similar. Error bars indicate statistical uncertainties only. }
\label{fig:v2_pPb}
\end{figure}

Figure~\ref{fig:v2_pPb} shows the $v_2(\etalab)/v_2(\etalab=0)$ and $v_3(\etalab)/v_3(\etalab=0)$ results obtained from the corresponding $V_2$ and $V_3$ data in Fig.~\ref{fig:V2_pPb}. The curves show the $v_n(\etalab)/v_n(\etalab=0)$ obtained from the high-multiplicity data alone, $V_n^\mathrm{HM}$, without subtraction of the low-multiplicity data. The data points are obtained from the low-multiplicity-subtracted $V_n^\text{sub}$; closed circles are from the Pb-side trigger particle data and open circles from the p-side. To avoid large contamination from short-range correlations, only the large $\abs{\deta}$ range is shown, but still with enough overlap in mid-rapidity $\etalab$ between the two trigger selections; good agreement is observed.
Significant pseudorapidity dependence is observed for the anisotropy parameter; it decreases by about $(24 \pm 4)\%$ (statistical uncertainty only) from $\etalab=0$ to $\etalab=2$ in the p-direction.
The behavior of the normalized $v_2(\etalab)/v_2(\etalab=0)$ is different in the Pb-side, with the maximum difference being smaller.
The $v_2$ appears to be asymmetric about $\etacm = 0$, which corresponds to $\etalab=0.465$. A non-zero $v_3$ is observed, however the uncertainties are too large to draw a definite conclusion regarding its pseudorapidity dependence.

When using long-range two-particle correlations to obtain anisotropic flow, the large pseudorapidity separation between the particles, while reducing nonflow effects, may lead to underestimation of  the anisotropic flow because of event plane decorrelation stemming from the fluctuating initial conditions~\cite{PhysRevC.87.011901,PhysRevC.91.044904}. This effect was studied in pPb and PbPb collisions~\cite{14012}. The observed decrease in $v_2$ with increasing absolute value of pseudorapidity could be partially due to such decorrelation.

\begin{figure}[thb]
\centering
\includegraphics[width=0.48\textwidth]{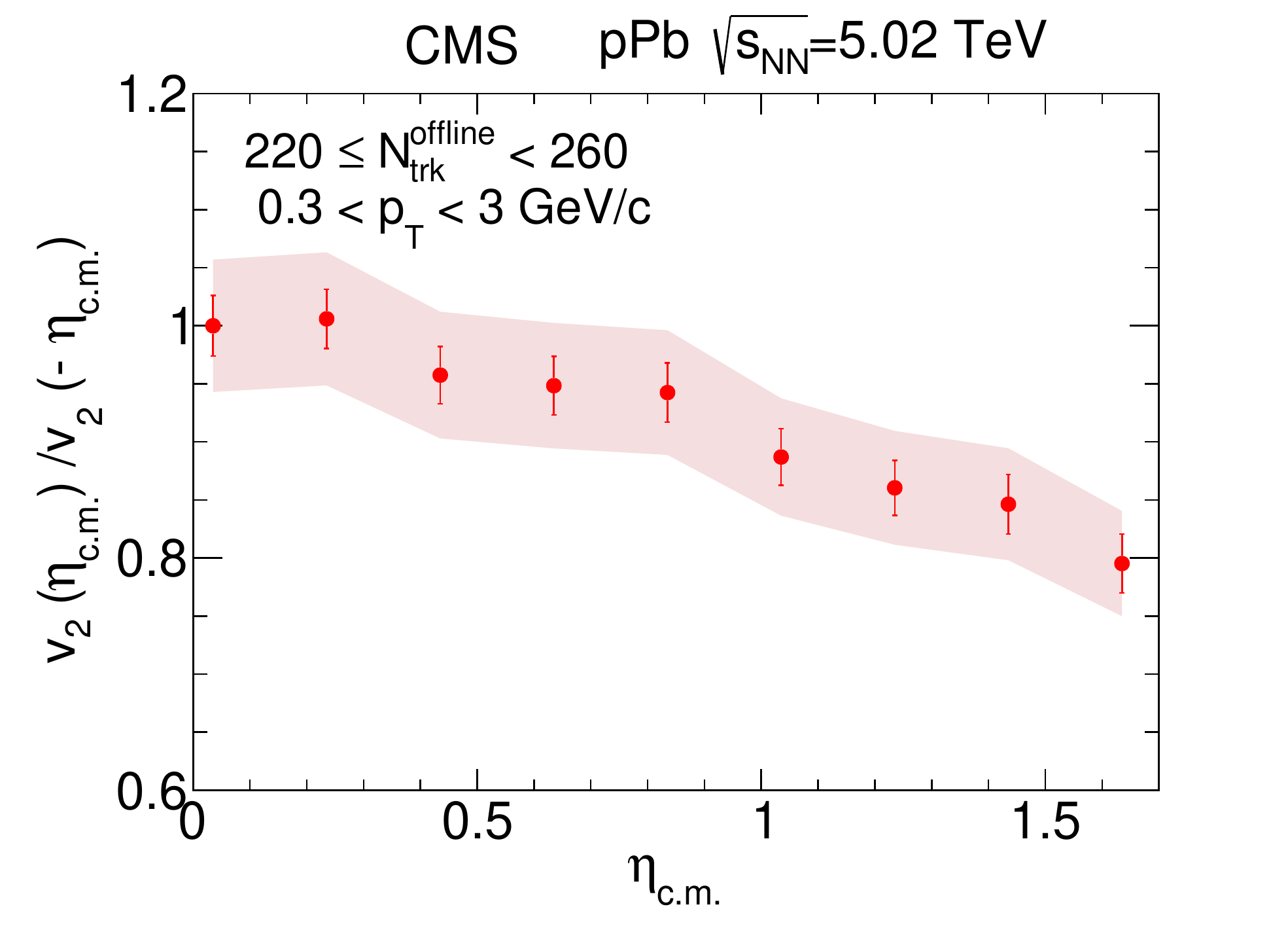}
\caption{(Color online) $v_2(\etacm)/v_2(-\etacm)$, as a function of $\etacm$ in the center-of-mass frame. The data points are results from $V_n^\text{sub}$ with low-multiplicity data subtracted. The bands show the systematic uncertainty of $\pm$5.7\%.
Error bars indicate statistical uncertainties only.}
\label{fig:v2_asym_ratio}
\end{figure}

The asymmetry of the azimuthal anisotropy is studied by taking the ratio of the $v_{n}$ value at positive $\etacm$ to the value at $-\etacm$ in the center-of-mass frame, as shown in Fig.~\ref{fig:v2_asym_ratio}.
The ratio shows a decreasing trend with increasing $\etacm$.

In pPb collisions, the average $\pt$ of charged hadrons depends on pseudorapidity. As stated in Ref.~\cite{Bozek:2014plb}, the pseudorapidity dependence of $\langle\pt\rangle$ could influence the pseudorapidity dependence of $v_2$. This may have relevance to the shape of the normalized $v_2$ distribution as observed in Fig.~\ref{fig:v2_pPb}.
To compare $v_2$ and the $\langle\pt\rangle$ distribution,
the $\pt$ spectra for different $\etacm$ ranges are obtained from Ref.~\cite{RpA12017}. The charged particle $\pt$ spectra in minimum-bias events are then fitted with a Tsallis function, as done in Ref.~\cite{10002}.

\begin{figure}[thb]
\centering
\includegraphics[width=0.48\textwidth]{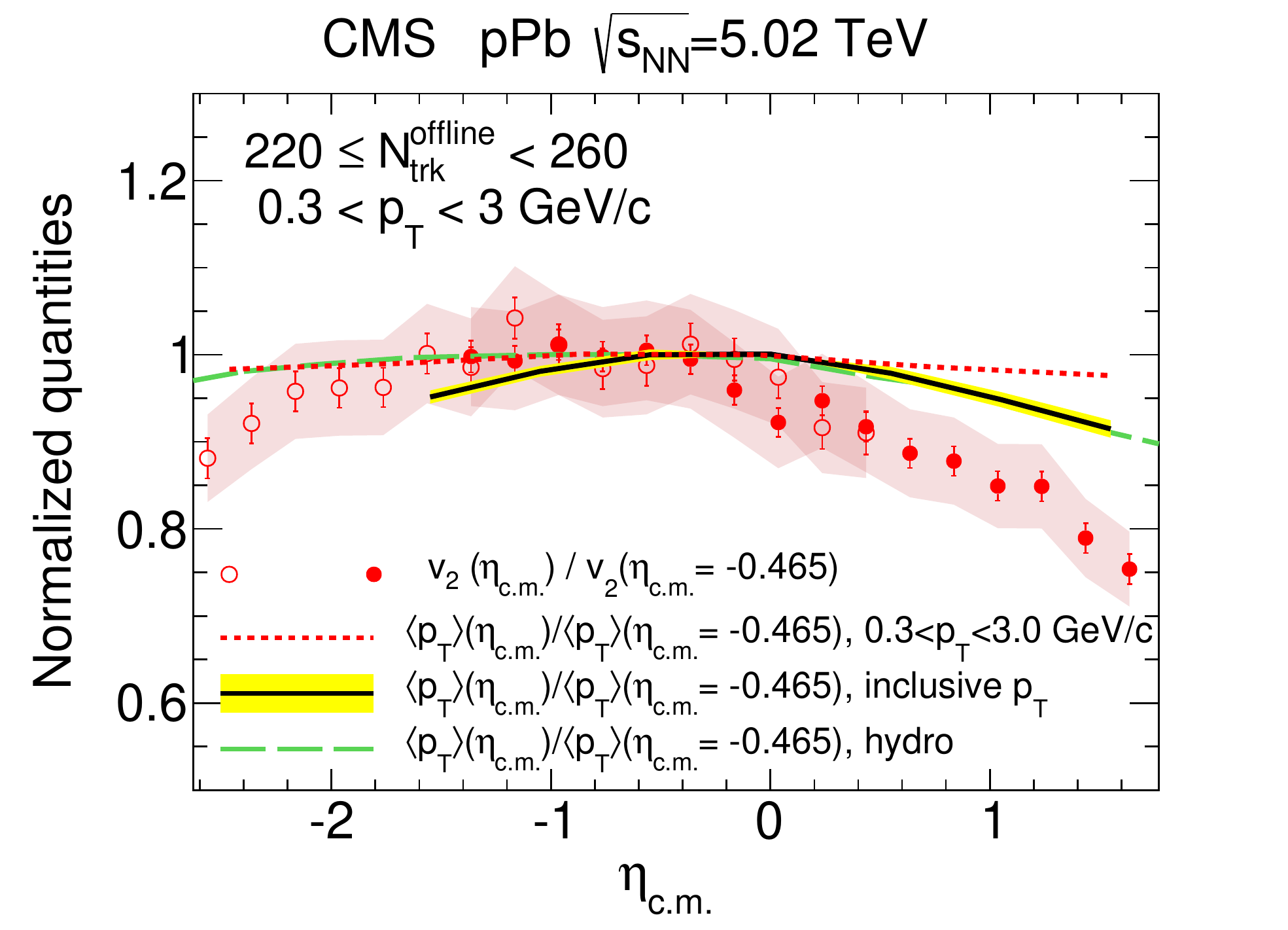}
\caption{(Color online) Self-normalized $v_2(\etacm)/v_2(\etacm=-0.465)$ distribution with low-multiplicity subtraction from Pb-side (filled circles) and p-side (open circles) triggers, and $\langle\pt\rangle(\etacm)/\langle\pt\rangle(\etacm = - 0.465)$ of $0< \pt<6 \GeVc$ range from minimum-bias events (solid line) and $0.3< \pt < 3\GeVc$ range from high-multiplicity ($220\leq\noff<260$) events (dotted line) as functions of $\etacm$. Dashed curve is the hydrodynamic prediction for $\langle\pt\rangle(\etacm)/\langle\pt\rangle(\etacm= -0.465)$ distribution.
\label{fig:v2_pt_compare}}
\end{figure}

The inclusive-particle $\pt$ is averaged within $0< \pt<6 \GeVc$. In addition, the average momentum for the particles used in this analysis, $0.3< \pt<3 \GeVc$ and $220\leq\noff<260$, is calculated and plotted in Fig.~\ref{fig:v2_pt_compare}.
The $\langle\pt\rangle$ as a function of $\etacm$ does not change for different multiplicity ranges within 1\%. Thus, the minimum bias $\langle\pt\rangle$ distribution is compared directly to the high-multiplicity anisotropy $v_2$ result.
The $\langle\pt\rangle$ distribution is normalized by its value at $\etacm$ = -0.465.
Self-normalized $\langle\pt\rangle(\etacm)/\langle\pt\rangle(\etacm= -0.465)$ is plotted on Fig.~\ref{fig:v2_pt_compare}, compared to the self-normalized $v_2(\etacm)/v_2(\etacm=-0.465)$ distribution in the center-of-mass frame. The systematic uncertainty band for $\langle\pt\rangle(\etacm)/\langle\pt\rangle(\etacm= -0.465)$ is obtained by averaging the upper and lower limits of the systematic uncertainty band from the underlying $\pt$ spectra.
The hydrodynamic theoretical prediction for $\langle\pt\rangle(\etacm)/\langle\pt\rangle(\etacm= -0.465)$ is also plotted.

As shown in Fig.~\ref{fig:v2_pt_compare}, the hydrodynamic calculation~\cite{Bozek:2014plb} for $\langle\pt\rangle$ falls more rapidly than the $\langle\pt\rangle$ for data (solid and dotted lines) towards positive $\etacm$. The distribution is asymmetric for both data and theory.
The comparison of the $\langle\pt\rangle$ and the $v_2$ distributions shows that both observables have a decreasing trend towards large $|\etacm|$, but the decrease in $\langle\pt\rangle$ at forward pseudorapidity is smaller. The decrease of $v_2$ with $\etacm$ does not appear to be entirely from a change in $\langle\pt\rangle$; other physics is likely at play. The value of $v_2$ decreases by $(20 \pm 4)\%$ (statistical uncertainty only) from $\etacm = 0$ to $\etacm\approx 1.5$.

\section{Summary}

Two-particle correlations as functions of $\dphi$ and $\deta$ are reported in \pPb\ collisions at $\rootsNN = 5.02$\TeV by the CMS experiment. The trigger particle is restricted to narrow pseudorapidity windows.
The combinatorial background is assumed to be uniform in $\dphi$ and normalized by the ZYAM procedure as a function of $\deta$. The near-side jet correlated yield is fitted and found to be greater in high-multiplicity than in low-multiplicity collisions.
The ridge yield is studied as a function of $\dphi$ and $\deta$ and it is found to depend on pseudorapidity and the underlying background shape ZYAM($\deta$). The pseudorapidity dependence differs for trigger particles selected on the proton and the Pb sides.

The Fourier coefficients of the two-particle correlations in high-multiplicity collisions are reported, with and without subtraction of the scaled low-multiplicity data. The pseudorapidity dependence of the single-particle anisotropy parameters, $v_2$ and $v_3$, is inferred.
Significant pseudorapidity dependence of $v_2$ is found. The distribution is asymmetric about $\etacm = 0$ with an approximate $(20 \pm 4)\%$ decrease from $\etacm = 0$ to $\etacm\approx 1.5$, and a smaller decrease towards the Pb-beam direction. Finite $v_3$ is observed, but the uncertainties are presently too large to draw conclusions regarding the pseudorapidity dependence.

{\tolerance=1200
The self-normalized $v_2(\etacm)/v_2(\etacm=-0.465)$ distribution is compared to the \breakhere$\langle\pt\rangle(\etacm)/\langle\pt\rangle(\etacm= -0.465)$ distribution as well as from hydrodynamic calculations.
The $\langle\pt\rangle(\etacm)/\langle\pt\rangle(\etacm= -0.465)$ distribution shows a decreasing trend towards positive $\etacm$. The $v_2(\etacm)/v_2(\etacm = -0.465)$ distribution also shows a decreasing trend towards positive $\etacm$, but the decrease is more significant in the case of the $v_2$ measurement. This indicates that physics mechanisms other than the change in the underlying particle
spectra, such as event plane decorrelation over pseudorapidity, may influence the anisotropic flow.
\par}

\begin{acknowledgments}
\hyphenation{Bundes-ministerium Forschungs-gemeinschaft Forschungs-zentren} We congratulate our colleagues in the CERN accelerator departments for the excellent performance of the LHC and thank the technical and administrative staffs at CERN and at other CMS institutes for their contributions to the success of the CMS effort. In addition, we gratefully acknowledge the computing centers and personnel of the Worldwide LHC Computing Grid for delivering so effectively the computing infrastructure essential to our analyses. Finally, we acknowledge the enduring support for the construction and operation of the LHC and the CMS detector provided by the following funding agencies: the Austrian Federal Ministry of Science, Research and Economy and the Austrian Science Fund; the Belgian Fonds de la Recherche Scientifique, and Fonds voor Wetenschappelijk Onderzoek; the Brazilian Funding Agencies (CNPq, CAPES, FAPERJ, and FAPESP); the Bulgarian Ministry of Education and Science; CERN; the Chinese Academy of Sciences, Ministry of Science and Technology, and National Natural Science Foundation of China; the Colombian Funding Agency (COLCIENCIAS); the Croatian Ministry of Science, Education and Sport, and the Croatian Science Foundation; the Research Promotion Foundation, Cyprus; the Ministry of Education and Research, Estonian Research Council via IUT23-4 and IUT23-6 and European Regional Development Fund, Estonia; the Academy of Finland, Finnish Ministry of Education and Culture, and Helsinki Institute of Physics; the Institut National de Physique Nucl\'eaire et de Physique des Particules~/~CNRS, and Commissariat \`a l'\'Energie Atomique et aux \'Energies Alternatives~/~CEA, France; the Bundesministerium f\"ur Bildung und Forschung, Deutsche Forschungsgemeinschaft, and Helmholtz-Gemeinschaft Deutscher Forschungszentren, Germany; the General Secretariat for Research and Technology, Greece; the National Scientific Research Foundation, and National Innovation Office, Hungary; the Department of Atomic Energy and the Department of Science and Technology, India; the Institute for Studies in Theoretical Physics and Mathematics, Iran; the Science Foundation, Ireland; the Istituto Nazionale di Fisica Nucleare, Italy; the Ministry of Science, ICT and Future Planning, and National Research Foundation (NRF), Republic of Korea; the Lithuanian Academy of Sciences; the Ministry of Education, and University of Malaya (Malaysia); the Mexican Funding Agencies (BUAP, CINVESTAV, CONACYT, LNS, SEP, and UASLP-FAI); the Ministry of Business, Innovation and Employment, New Zealand; the Pakistan Atomic Energy Commission; the Ministry of Science and Higher Education and the National Science Centre, Poland; the Funda\c{c}\~ao para a Ci\^encia e a Tecnologia, Portugal; JINR, Dubna; the Ministry of Education and Science of the Russian Federation, the Federal Agency of Atomic Energy of the Russian Federation, Russian Academy of Sciences, and the Russian Foundation for Basic Research; the Ministry of Education, Science and Technological Development of Serbia; the Secretar\'{\i}a de Estado de Investigaci\'on, Desarrollo e Innovaci\'on and Programa Consolider-Ingenio 2010, Spain; the Swiss Funding Agencies (ETH Board, ETH Zurich, PSI, SNF, UniZH, Canton Zurich, and SER); the Ministry of Science and Technology, Taipei; the Thailand Center of Excellence in Physics, the Institute for the Promotion of Teaching Science and Technology of Thailand, Special Task Force for Activating Research and the National Science and Technology Development Agency of Thailand; the Scientific and Technical Research Council of Turkey, and Turkish Atomic Energy Authority; the National Academy of Sciences of Ukraine, and State Fund for Fundamental Researches, Ukraine; the Science and Technology Facilities Council, UK; the US Department of Energy, and the US National Science Foundation.

Individuals have received support from the Marie-Curie program and the European Research Council and EPLANET (European Union); the Leventis Foundation; the A. P. Sloan Foundation; the Alexander von Humboldt Foundation; the Belgian Federal Science Policy Office; the Fonds pour la Formation \`a la Recherche dans l'Industrie et dans l'Agriculture (FRIA-Belgium); the Agentschap voor Innovatie door Wetenschap en Technologie (IWT-Belgium); the Ministry of Education, Youth and Sports (MEYS) of the Czech Republic; the Council of Science and Industrial Research, India; the HOMING PLUS program of the Foundation for Polish Science, cofinanced from European Union, Regional Development Fund; the Mobility Plus program of the Ministry of Science and Higher Education (Poland); the OPUS program of the National Science Center (Poland); MIUR project 20108T4XTM (Italy); the Thalis and Aristeia programs cofinanced by EU-ESF and the Greek NSRF; the National Priorities Research Program by Qatar National Research Fund; the Rachadapisek Sompot Fund for Postdoctoral Fellowship, Chulalongkorn University (Thailand); the Chulalongkorn Academic into Its 2nd Century Project Advancement Project (Thailand); and the Welch Foundation, contract C-1845.
\end{acknowledgments}

\bibliography{auto_generated}

\cleardoublepage \appendix\section{The CMS Collaboration \label{app:collab}}\begin{sloppypar}\hyphenpenalty=5000\widowpenalty=500\clubpenalty=5000\textbf{Yerevan Physics Institute,  Yerevan,  Armenia}\\*[0pt]
V.~Khachatryan, A.M.~Sirunyan, A.~Tumasyan
\vskip\cmsinstskip
\textbf{Institut f\"{u}r Hochenergiephysik der OeAW,  Wien,  Austria}\\*[0pt]
W.~Adam, E.~Asilar, T.~Bergauer, J.~Brandstetter, E.~Brondolin, M.~Dragicevic, J.~Er\"{o}, M.~Flechl, M.~Friedl, R.~Fr\"{u}hwirth\cmsAuthorMark{1}, V.M.~Ghete, C.~Hartl, N.~H\"{o}rmann, J.~Hrubec, M.~Jeitler\cmsAuthorMark{1}, V.~Kn\"{u}nz, A.~K\"{o}nig, M.~Krammer\cmsAuthorMark{1}, I.~Kr\"{a}tschmer, D.~Liko, T.~Matsushita, I.~Mikulec, D.~Rabady\cmsAuthorMark{2}, N.~Rad, B.~Rahbaran, H.~Rohringer, J.~Schieck\cmsAuthorMark{1}, R.~Sch\"{o}fbeck, J.~Strauss, W.~Treberer-Treberspurg, W.~Waltenberger, C.-E.~Wulz\cmsAuthorMark{1}
% ...
\vskip\cmsinstskip
% ...
1:~~Also at Vienna University of Technology, Vienna, Austria\\
2:~~Also at CERN, European Organization for Nuclear Research, Geneva, Switzerland\\
% ...

\end{sloppypar}
\end{document}